\documentclass[journal]{IEEEtran}
\makeatletter

\usepackage{blindtext}
\usepackage[T1]{fontenc}
\usepackage{textcomp}
\usepackage{amsmath}
\usepackage{amssymb}
\usepackage{bm}
\usepackage{mdwmath}
\usepackage{cases}
\usepackage[short, c2]{optidef}
\usepackage{graphicx}
\graphicspath{{Figure/PDF/}{Biography/PDF/}}
\usepackage[dvipsnames]{xcolor}
\definecolor{myblue}{rgb}{0,0.4980,1} 
\definecolor{myred}{rgb}{0.8706,0.1608,0.0627} 

\usepackage[caption=false,font=footnotesize,subrefformat=parens]{subfig}

\usepackage[nosort,noadjust]{cite}
\usepackage{url}
\usepackage{tabularray}
\UseTblrLibrary{varwidth}
\UseTblrLibrary{diagbox}
\ExplSyntaxOn
\tl_const:Nn \c__tblr_valign_M_tl {M}
\tl_new:N    \g__tblr_cell_valign_sub_tl

\keys_define:nn { tblr-column }
  {
    M .meta:n = { valign = M },
  }

\cs_gset_protected:Npn \__tblr_get_cell_alignments:nn #1 #2
  {
    \group_begin:
    \tl_gset:Nx \g__tblr_cell_halign_tl
      { \__tblr_data_item:neen { cell } {#1} {#2} { halign } }
    \tl_set:Nx \l__tblr_v_tl
      { \__tblr_data_item:neen { cell } {#1} {#2} { valign } }
    \tl_case:NnF \l__tblr_v_tl
      {
        \c__tblr_valign_t_tl
          {
            \tl_gset:Nn \g__tblr_cell_valign_tl {m}
            \tl_gset:Nn \g__tblr_cell_middle_tl {t}
            \tl_gclear:N \g__tblr_cell_valign_sub_tl
          }
        \c__tblr_valign_m_tl
          {
            \tl_gset:Nn \g__tblr_cell_valign_tl {m}
            \tl_gset:Nn \g__tblr_cell_middle_tl {m}
            \tl_gclear:N \g__tblr_cell_valign_sub_tl
          }
        \c__tblr_valign_M_tl
          {
            \tl_gset:Nn \g__tblr_cell_valign_tl {m}
            \tl_gset:Nn \g__tblr_cell_valign_sub_tl {M}
            \tl_gset:Nn \g__tblr_cell_middle_tl {t}
          }
        \c__tblr_valign_b_tl
          {
            \tl_gset:Nn \g__tblr_cell_valign_tl {m}
            \tl_gset:Nn \g__tblr_cell_middle_tl {b}
            \tl_gclear:N \g__tblr_cell_valign_sub_tl
          }
      }
      {
        \tl_gset_eq:NN \g__tblr_cell_valign_tl \l__tblr_v_tl
        \tl_gclear:N \g__tblr_cell_middle_tl
      }
    \group_end:
  }

\cs_gset_protected:Npn \__tblr_build_cell_content:NN #1 #2
  {
    \hbox_set_to_wd:Nnn \l__tblr_a_box { \l__tblr_cell_wd_dim }
      {
        \tl_if_eq:NnTF \g__tblr_cell_halign_tl {j}
          { \__tblr_get_cell_text:nn {#1} {#2} \hfil }
          {
            \tl_if_eq:NnF \g__tblr_cell_halign_tl {l} { \hfil }
            \__tblr_get_cell_text:nn {#1} {#2}
            \tl_if_eq:NnF \g__tblr_cell_halign_tl {r} { \hfil }
          }
      }
    \vbox_set_to_ht:Nnn \l__tblr_b_box { \l__tblr_cell_ht_dim }
      {
        \tl_case:Nn \g__tblr_cell_valign_tl
          {
            \c__tblr_valign_m_tl
              {
                \vfil
                \bool_lazy_and:nnT
                  { \int_compare_p:nNn { \l__tblr_cell_rowspan_tl } < {2} }
                  { !\tl_if_eq_p:NN \c__tblr_valign_M_tl \g__tblr_cell_valign_sub_tl }
                  {
                    \box_set_ht:Nn \l__tblr_a_box
                      { \__tblr_data_item:nen { row } {#1} { @row-upper } }
                    \box_set_dp:Nn \l__tblr_a_box
                      { \__tblr_data_item:nen { row } {#1} { @row-lower } }
                  }
                \box_use:N \l__tblr_a_box
                \vfil
              }
            \c__tblr_valign_h_tl
              {
                \box_set_ht:Nn \l__tblr_a_box
                  { \__tblr_data_item:nen { row } {#1} { @row-head } }
                \box_use:N \l__tblr_a_box
                \vfil
              }
            \c__tblr_valign_f_tl
              {
                \vfil
                \int_compare:nNnTF { \l__tblr_cell_rowspan_tl } < {2}
                  {
                    \box_set_dp:Nn \l__tblr_a_box
                      { \__tblr_data_item:nen { row } {#1} { @row-foot } }
                  }
                  {
                    \box_set_dp:Nn \l__tblr_a_box
                      {
                        \__tblr_data_item:nen
                          { row }
                          { \int_eval:n { #1 + \l__tblr_cell_rowspan_tl - 1 } }
                          { @row-foot }
                      }
                  }
                \box_use:N \l__tblr_a_box
              }
          }
        \hrule height ~ 0pt 
      }
    \vbox_set_to_ht:Nnn \l__tblr_c_box
      { \l__tblr_row_ht_dim - \l__tblr_row_abovesep_dim }
      {
        \box_use:N \l__tblr_b_box
        \vss
      }
    \skip_horizontal:n { \l__tblr_x_tl }
    \box_use:N \l__tblr_c_box
    \skip_horizontal:n { \l__tblr_y_tl - \l__tblr_cell_wd_dim + \l__tblr_w_tl }
  }
\ExplSyntaxOff
\usepackage{pifont}
\UseTblrLibrary{counter}
\usepackage[linesnumbered,ruled]{algorithm2e}

\usepackage{hyperref}
\newcommand{\colorhypersetup}{\@ifpackageloaded{hyperref}{\hypersetup{%
	bookmarksopen=true,%
	bookmarksnumbered=true,%
	pdfpagemode={UseOutlines},
	pdfstartview={FitH},%
	colorlinks=true,%
	linkcolor={myred},%
	citecolor={orange}
}}{\empty}}
\newcommand{\blackhypersetup}{\@ifpackageloaded{hyperref}{\hypersetup{%
    colorlinks=true,%
    linkcolor=.,%
    citecolor=.,%
    filecolor=.,%
    urlcolor=.
}}{\empty}}
 \blackhypersetup

\usepackage{mfirstuc-english}
\MFUhyphentrue
\usepackage[version3,description,IEEEtran]{eacro}
\acsetup{%
    pages/display=all,%
    pages/seq/use=false,%
    pages/name=true,%
    pages/fill={\quad},%
    make-links=true%
}

\DeclareAcronym{bs}{
	short = BS,
	long = base station}
 \DeclareAcronym{es}{
	short = ES,
	long = edge server}
 \DeclareAcronym{mg}{
	short = MG,
	long = multicast group}
  \DeclareAcronym{msvs}{
	short = MSVS,
	long = multicast short video streaming}
 \DeclareAcronym{ai}{
	short = AI,
	long = artificial intelligence}
  \DeclareAcronym{dt}{
	short = DT,
	long = digital twin}
 \DeclareAcronym{andcm}{
	short = ADT4AM,
	long = AI-native network DT architecture for autonomous network management}
 \DeclareAcronym{udt}{
	short = UDT,
	long = user DT}
 \DeclareAcronym{idt}{
	short = IDT,
	long = infrastructure DT}
 \DeclareAcronym{sdt}{
	short = SDT,
	long = slice DT}
 \DeclareAcronym{nc}{
	short = NC,
	long = network controller}
\DeclareAcronym{cnn}{
	short = CNN,
	long = convolutional neural network}
\DeclareAcronym{rnn}{
	short = RNN,
	long = recurrent neural network}
 \DeclareAcronym{gan}{
	short = GAN,
	long = generative adversarial network}
  \DeclareAcronym{nlp}{
	short = NLP,
	long = natural language processing}
 \DeclareAcronym{bleu}{
	short = BLEU,
	long = bilingual evaluation understudy}
 \DeclareAcronym{rouge}{
	short = ROUGE,
	long = recall-oriented understudy for gisting evaluation}
 \DeclareAcronym{gnn}{
	short = GNN,
	long = graph neural network}
  \DeclareAcronym{drl}{
	short = DRL,
	long = deep reinforcement learning}
 \DeclareAcronym{llm}{
	short = LLM,
	long = large language model}
  \DeclareAcronym{ap}{
	short = AP,
	long = access point}
   \DeclareAcronym{iot}{
	short = IoT,
	long = Internet of things}
 \DeclareAcronym{qoe}{
	short = QoE,
	long = quality of experience}
 \DeclareAcronym{lstm}{
	short = LSTM,
	long = long short-term memory}
 \DeclareAcronym{ddqn}{
	short = DDQN,
	long = double deep Q network}
  \DeclareAcronym{cs}{
	short = CS,
	long = cloud server}
 \DeclareAcronym{svm}{
	short = SVM,
	long = support vector machine}
 \DeclareAcronym{knn}{
	short = K-NN,
	long = K-nearest neighbors}

\newcommand{\upperroman}[1]{\uppercase\expandafter{\romannumeral#1}}

\newcommand{\myunit}[1]{%
	\ifmmode
		\mathrm{#1}
	\else
		$ \mathrm{#1} $
	\fi}
\newcommand{\murm}{%
	\ifmmode
		\text{\textmu}
	\else
		\textmu
	\fi}

\newcommand{\MYnewpage}{%
	\ifCLASSOPTIONonecolumn
		\ifCLASSOPTIONjournal
			\typeout{The onecolumn journal mode.}
			\newpage
		\fi
	\fi}

\newlength{\mysinglefigwidth}
\newlength{\mymultifigwidth}
\ifCLASSOPTIONonecolumn
	\AtBeginDocument{\setlength{\mysinglefigwidth}{0.7\linewidth}}
\else
	\AtBeginDocument{\setlength{\mysinglefigwidth}{\linewidth}}
\fi
\ifCLASSOPTIONonecolumn
	\AtBeginDocument{\setlength{\mymultifigwidth}{0.5\linewidth}}
\else
	\AtBeginDocument{\setlength{\mymultifigwidth}{\linewidth}}
\fi

\makeatother
\begin{document}
\ifCLASSOPTIONonecolumn
    \typeout{The onecolumn mode.}
    \title{\LARGE Title}
\author{\small Wen Wu,~\IEEEmembership{\small Senior~Member,~IEEE,}
	and
	Xuemin~(Sherman)~Shen,~\IEEEmembership{\small Fellow,~IEEE}
	\thanks{W. Wu is with the Frontier Research Center, Peng Cheng Laboratory, Shenzhen, Guangdong,  China, 518055 (email: wuw02@pcl.ac.cn);}
	\thanks{X. Shen are with the Department of Electrical and Computer Engineering, University of Waterloo, Waterloo, Ontario, Canada, N2L 3G1  (email:  sshen@uwaterloo.ca); }
	
}
\else
    \typeout{The twocolumn mode.}
    \title{AI-Native Network Digital Twin for Intelligent Network Management in 6G}
\author{\small Wen Wu,~\IEEEmembership{\small Senior~Member,~IEEE,}
	Xinyu Huang,~\IEEEmembership{\small Student Member,~IEEE,}
	and 
	Tom H. Luan,~\IEEEmembership{\small Senior Member,~IEEE}
	\thanks{W. Wu is with the Frontier Research Center, Peng Cheng Laboratory, Shenzhen, Guangdong,  China (email: wuw02@pcl.ac.cn);}
	\thanks{X. Huang is with the Department of Electrical and Computer Engineering, University of Waterloo, Waterloo, Ontario, Canada (email:  xinyu.huang1@uwaterloo.ca); }
	\thanks{T. H. Luan is with the School of Cyber Science and Engineering, Xi’an Jiaotong University, Xi’an, China (Email: tom.luan@xjtu.edu.cn)}
	
}
    
\fi

\ifCLASSOPTIONonecolumn
	\typeout{The onecolumn mode.}
\else
	\typeout{The twocolumn mode.}
\fi

\maketitle

\ifCLASSOPTIONonecolumn
	\typeout{The onecolumn mode.}
	\vspace*{-50pt}
\else
	\typeout{The twocolumn mode.}
\fi
\begin{abstract}
As a pivotal virtualization technology, network digital twin is expected to accurately reflect real-time status and abstract  features in the on-going sixth generation (6G) networks. 
In this article, we propose an  artificial intelligence (AI)-native network digital twin framework for 6G networks to enable the synergy of AI and network digital twin, thereby facilitating intelligent network management. In the proposed framework, AI models are utilized to establish network digital twin models to facilitate network status prediction, network pattern abstraction, and network management decision-making. Furthermore, 
potential solutions are proposed for enhance the performance of network digital twin. Finally, a case study is presented, followed by a discussion of open research issues that are essential for AI-native network digital twin in 6G networks.
\end{abstract}

\ifCLASSOPTIONonecolumn
	\typeout{The onecolumn mode.}
	\vspace*{-10pt}
\else
	\typeout{The twocolumn mode.}
\fi

\IEEEpeerreviewmaketitle

\MYnewpage


\section{Introduction}
\label{sec:Introduction}

With the rollout of the fifth generation (5G) networks across the world, industry and academia have started looking into the next generation wireless networks, i.e., 6G. As 5G's successor, 6G is expected to support unprecedentedly diverse services with substantially increased service demands and evolve towards intelligent networking~\cite{wu2021ai}. \emph{Artificial intelligence} (AI), as anticipated, will penetrate every corner of the network, including end users, the network edge, and the cloud, resulting in network intelligence. The intelligence, spreading over network-wide entities, would facilitate intelligent network management for the on-going 6G networks.

Digital twin is another potential technology to put forward advanced network management in 6G networks~\cite{holi}. The concept of digital twin originates from the manufacturing industry~\cite{grieves2014digital}, which constructs detailed digital replicas of physical systems to simulate, monitor, and optimize the physical systems in real-time, thereby providing a significant leap in operational efficiency. This concept has been progressively applied to the wireless networks, i.e., \emph{network digital twin}. Network digital twin enhances traditional network management performance by integrating three key components: user digital twin, infrastructure digital twin, and slice digital twin. User digital twins capture and model individual user behavior patterns, thereby supporting the analysis of personalized service demands for network optimization. Infrastructure digital twin correspond to network infrastructures, such as base stations, routers and switches, edge servers, and cloud servers. Infrastructure digital twins are instrumental in analyzing network traffic patterns, balancing network workload, and validating network management policies to ensure the robustness and efficiency of mobile communication networks. Slice digital twins are essential for managing network slices to enhance user \ac{qoe}. The relationships between historical service requests, user QoE, and slicing policies can be learned by the data analysis models in slice digital twin. Through ongoing cross-validation within the virtual networks, slice digital twins develop tailored slicing strategies to improve user QoE. These digital twins contribute to a comprehensive network management framework through effective status simulation, predictive data analytics, and reliable policy validation, thereby enhancing network performance.


{However, achieving AI-native network digital twin faces the following challenges. Firstly, collecting massive data for network digital twin construction places high traffic pressure on communication networks. Therefore, developing efficient data transmission techniques is essential to reduce data collection overhead. Secondly, the spatial-temporal dynamics of communication networks require a scalable network digital twin model to conduct low-energy data processing. Therefore, it is crucial to develop a scalable network digital twin model inference mechanism to enhance energy efficiency. Thirdly, network digital twin can be gradually inaccurate if network digital twin models cannot adapt to inherent data drift within the network, triggered by changes in user behaviors, device interoperability, and service demands. Therefore, it is essential to develop an adaptive network digital twin model update method to improve network digital twin fidelity.
	
\begin{figure*}[t]
	\centering
	\includegraphics[width=0.85\textwidth]{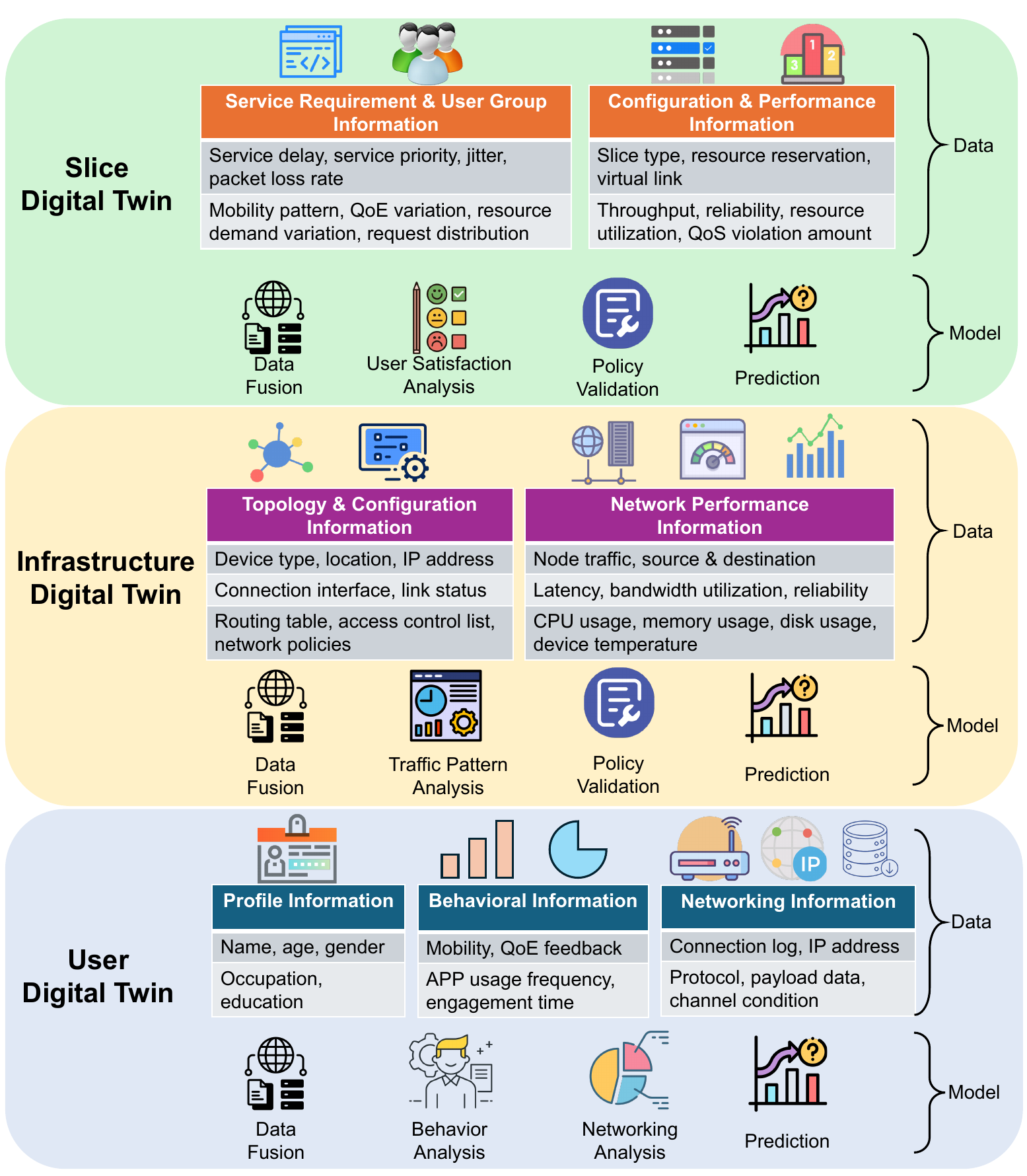}
	\caption{The three categories of network digital twin, i.e., user, infrastructure, and slice digital twins.}
	\label{dt_class}
\end{figure*}

{In this article, we propose an AI-native network digital twin framework for facilitating intelligent network management in 6G.  AI models are cooperatively embedded in user, infrastructure, and slice digital twins to realize AI-native network status emulation, network feature abstraction, and network decision-making, respectively.} To address the above challenges, we first develop an adaptive network digital twin model inference scheme to reduce energy consumption. Next, we propose a dual error-based model update mechanism to adaptively trigger network digital twin model updates in an online manner. A case study pertaining to AI-native adaptive user digital twin model inference and update is presented, followed by a discussion on research issues.


\section{Network Digital Twin and AI Technology} \label{sec. Digital Twin Architecture}

\subsection{Network Digital Twin}
Network digital twin is a virtual representation that accurately replicates and simulates the structure, operations, and dynamics of a physical communication network. Network digital twin usually consists of network status data and various specialized models. The former is used to reflect real-time network status, while the latter is utilized to analyze network characteristics, predict network status, validate network management policies, and optimize network performance, etc. Based on different types of virtualized physical network entities, network digital twin consists of three categories: user digital twin, infrastructure digital twin, and slice digital twin, as shown in Fig.~\ref{dt_class}, which are detailed as follows.

\subsubsection{User Digital Twin}
One user digital twin corresponds to an end user that characterizes its fine-grained user status. The stored user status data are classified into three types, i.e., profile information, behavioral information, and networking information. Initial user digital twin's data are collected from the user’s devices, while subsequent data can be generated through the prediction model. {User digital twins are usually deployed at network edge nodes, which can help local network controllers perceive real-time user status and make timely network management decisions.} {In the user digital twin, the data are usually stored in JSON, XML, or CSV formats for readability and widespread compatibility with various systems.} To extract data from various sources, transform them into a unified format, and load them into a central data warehouse, the  user digital twin requires efficient data transformation and integration models. A key advantage of  user digital twin is that it can consolidate behavioral and networking information at a single network node for efficient correlation analysis, which enables intelligent network management. 

\subsubsection{Infrastructure Digital Twin}
One infrastructure digital twin is a digital mirror of a network infrastructure, which reflects its network status, configuration, and performance information. The information can be obtained from real-time reports of network infrastructures or prediction models. {Infrastructure digital twins are usually deployed at network edge nodes, which can help local network controller perceive real-time network infrastructure status to enable efficient communication, sensing, and computing collaboration.} The  infrastructure digital twin consists of extensive network data and sophisticated modeling techniques to optimize network management. The detailed topology and configuration information (in JSON/XML format) alongside performance metrics (in PCAP/CSV format) are collected and integrated from the network infrastructure. The data are processed via advanced methods, including data transformation and integration, which streamline disparate data into a cohesive framework, and traffic pattern analysis, which alleviates network overloads on certain network infrastructures. Furthermore, the  infrastructure digital twin employs a simulated network environment to validate that network management decisions can achieve expected performance.

\subsubsection{Slice Digital Twin}
A slice digital twin corresponds to a network slice, i.e., one isolated logical network in a partitioned physical network, which is responsible for managing the virtualized resources to satisfy different service requirements~\cite{9247519}. slice digital twin usually store data related to service requirement and user group information as well as configuration and performance information, which can be abstracted from  user digital twin and  infrastructure digital twin. Slice digital twins are usually deployed at network core nodes, which can help global network controller efficiently slice the whole network resources. The slice digital twin consists of two primary components: data and models. The data component encompasses detailed service requirements, such as delay and priority, alongside user group information like mobility patterns and resource demands, all formatted in JSON and XML format. {Slicing configuration and performance data are also stored in the slice digital twin, including slice type, resource reservation, and QoS violation metrics in JSON, XML, and CSV formats.} Regarding models, the slice digital twin utilizes data transformation and integration to ensure data consistency across various sources. {The slice digital twin adopts user satisfaction analysis to evaluate the slicing performance for adjusting dynamic slicing polices.} Moreover, policy validation ensures that operational policies align with compliance standards and performance objectives, while predictive models can facilitate proactive network resource reservation. Collectively, these elements can help slice digital twin accurately adjust network slicing strategies to satisfy evolving user requirements.


\subsection{AI Technology}
Three types of \ac{ai} technologies, i.e., machine learning, deep learning, and \ac{nlp}, are selected to explore the potential applications in the field of network digital twin, as shown in Table~\ref{ai}.

\subsubsection{Machine Learning}
As a transformative branch of \ac{ai} technologies, machine learning algorithms leverages its lightweight techniques, including \ac{svm}, random forest, and \ac{knn}, to enhance the predictive maintenance capabilities and operational efficiencies of network digital twin~\cite{9711524}. By implementing \ac{svm}, network digital twin can efficiently handle classification tasks, such as identifying different types of network traffic or detecting anomalies. The kernel function and margin maximization enable accurate predictions even in complex network environments. Random forest can improve reliability in predicting network performance and detecting potential failures by analyzing multiple decision trees. Furthermore, \ac{knn} can be utilized within the network digital twin  for tasks that require real-time decision-making based on the closest historical data points, such as real-time traffic classification and immediate load balancing decisions. Each learning paradigm has specific cache and computational requirements, typically ranging from low to medium, to implement efficient model deployment and inference.
\begin{table*}[!t]
	\renewcommand{\IEEEiedlistdecl}{\setlength{\IEEElabelindent}{0pt}}
	\centering
	\caption{The comparison of machine learning, deep learning, and NLP algorithms.}
	\label{ai}
	\begin{tblr}{
			width = \linewidth,
			colspec = {X[0.27,c,M]X[0.77,l,m]X[0.77,l,m]X[0.8,l,m]},
			hlines,
			hline{2} = {1}{-}{},
			vline{2-4},
			row{1} = {c,font=\bfseries},
			column{1} = {font=\bfseries},       
			columns = {rightsep=3pt},
			cell{3}{4} = {l},
			cell{2}{5} = {c},
			measure=vbox,
		}
		\textbf{Category} & \textbf{Machine Learning} & \textbf{Deep Learning} & \textbf{Natural Language Processing} \\
		\hline
		\textbf{Typical Algorithms} & 
		- Support vector machine \newline 
		- Random forest \newline 
		- K-nearest neighbors & 
		- Convolutional neural network \newline 
		- Autoencoder \newline 
		- Generative adversarial network & 
		- Word embedding \newline 
		- Recurrent neural network \newline 
		- Large language model \\
		\textbf{Characteristics} & 
		- Kernel function \& margin maximization \newline 
		- Decision trees \& overfitting resistance \newline 
		- Instance-based \& parameter dependency & 
		- Local connectivity \& pooling layers \newline 
		- Encoder \& decoder \newline 
		- Generator \& discriminator & 
		- Dense representation \& semantic similarity \newline 
		- Sequential processing \& variable length \newline 
		- Transformer \& attention mechanism \\
		\textbf{Functions} & 
		- Classification \& regression & 
		- Feature extraction \& data generation& 
		- Representation \& understanding\\
		\textbf{Memory} & 
		- Low to medium  & 
		- Medium to high & 
		- Medium to high \\
		
		\textbf{Computation} & 
		- Low to medium & 
		- Medium to high & 
		- High \\
		
	\end{tblr}
\end{table*}

\subsubsection{Deep Learning}
The deep learning algorithms leverages a multi-layered framework, such as \acp{cnn}, autoencoders, and \acp{gan}, to process network data with complex structures. Specifically, characterized by its ability to learn local connectivity patterns through pooling and convolutional layers, deep learning excels in handling high-dimensional data and generating or distinguishing between real and synthetic data through adversarial processes. Deep learning can conduct efficient extraction of intricate features from vast amounts of network data, significantly improving the accuracy and efficiency of predictive analytics in network digital twin~\cite{9508904}. For instance, \acp{cnn} can analyze spatial patterns for anomaly detection in network traffic, while autoencoders are adept at compressing complex and high-dimensional network data. Additionally, \acp{gan} can be employed to simulate network environments for stress testing without the risk of disrupting actual network operations. The integration of deep learning into network digital twin not only optimizes the performance and reliability of communication networks but also supports proactive maintenance strategies.

\subsubsection{Natural Language Processing}
The \ac{nlp} algorithms employ a variety of techniques designed for enabling computers to understand and manipulate human language, making it pivotal for constructing and maintaining network digital twin. \ac{nlp} algorithms include word embedding that provides dense representations of words capturing semantic similarities in network digital twin data; \acp{rnn} that excel in processing sequential network digital twin data; and transformers equipped with attention mechanisms that manage variable-length inputs and prioritize contextually important aspects of network digital twin data. Performance metrics, such as cosine similarity, perplexity, \ac{bleu}, and \ac{rouge} scores~\cite{dong2022survey}, help evaluate the effectiveness of network digital twin models. For the network digital twin, \ac{nlp} can significantly elevate operational intelligence and user interaction. For instance, by analyzing maintenance logs, user feedback, and other textual data, network digital twin can gain insights from historical network performance and adjust network management strategies. Techniques such as sentiment analysis and text summarization can automate user satisfaction monitoring, which can facilitate proactive network management~\cite{9937052, 10487933}. The integration of \ac{nlp} not only streamlines operational processes but also enhances the adaptability and responsiveness of network digital twin.

\section{AI-Native Network Digital Twin Framework for Intelligent Network Management}\label{sec: framework}

\subsection{Framework Overview}
{As shown in Fig.~\ref{framework}, we propose the overall AI-native network digital twin framework for intelligent network management, which can utilize the advantage of AI technology in data generation, data analysis, and decision-making.} The proposed framework consists of physical network, user digital twin, infrastructure digital twin, and slice digital twin. The physical network is a multi-tier network framework, where the centralized controller is responsible for large-timescale network slicing and the local controller handles the real-time network operations. Data from  user digital twins are initially abstracted into infrastructure digital twin, where they are merged with infrastructure status to facilitate real-time network operations. Subsequently, data from both  user digital twin and infrastructure digital twin are aggregated into slice digital twin to refresh slice status and refine network slicing strategies. In the following, we will show how to leverage the framework to facilitate intelligent network management. 

\begin{figure*}[!t]
	\centering  \includegraphics[width=2\mysinglefigwidth]{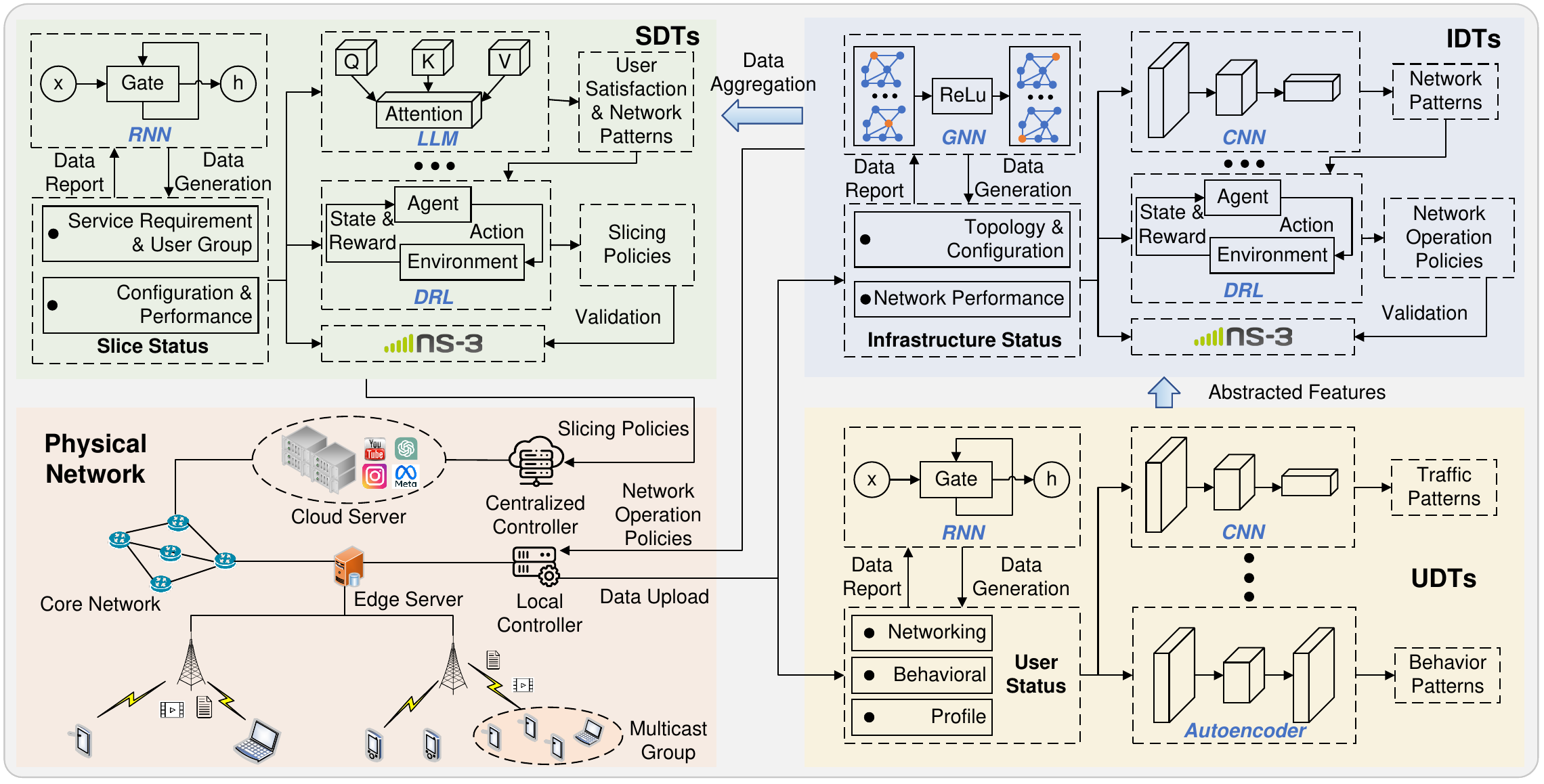}
	\caption{The proposed AI-native network digital twin framework for intelligent network management.}
	\label{framework}
\end{figure*}

\subsection{AI-Native User Digital Twin}
{Since an essential function of user digital twin} is to characterize real-time user status on the network side, frequently collecting data from the physical network not only incurs high costs but also poses risks of communication interruptions due to user mobility. \acp{rnn}, known for their proficiency in analyzing sequential and time-series data, are pivotal in continuously updating user status by capturing temporal patterns. By deploying lightweight \acp{rnn} models on edge servers, user status in user digital twin can be timely updated by data generation based on historical data. Another critical function of  user digital twin is abstracting users' traffic and behavior patterns. \acp{cnn} can effectively analyze network conditions to extract complex spatial and temporal patterns. Furthermore, autoencoders streamline the complexity of  user digital twin by distilling high-dimensional data into its most salient features through the encoding-decoding mechanism. This process transforms the intricate  user digital twin data into useful features, such as behavior patterns, thereby boosting the network management efficiency. Together, these \ac{ai} models can help  user digital twin monitor real-time user status, abstract behavior patterns, and compress high-dimensional user data into low-dimensional representation.} 

\subsection{AI-Native Infrastructure Digital Twin}
{As shown in Fig.~\ref{framework}, \acp{gnn}, \acp{cnn}, \ac{drl} models, and the network simulator, such as NS-3, are integrated into infrastructure digital twin to enhance the efficiency of construction and maintenance. GNNs are utilized to analyze the network's topology and configuration. They excel in handling the correlated data inherent in networks, such as connections between network infrastructures and the transmission of signals or data through these connections~\cite{10118812}. CNNs are adept at recognizing patterns from multi-dimensional data, including traffic flow, bandwidth usage, or anomaly detection. DRL employs an agent-based approach where each agent interacts with a simulated network environment provided by the network simulator, optimizing network operations through continuous learning from actions and received rewards~\cite{peng2020multi}. The network simulator serves as a pivotal tool in the infrastructure digital twin by providing a realistic simulation environment for network policy evaluation, ensuring that infrastructure digital twin accurately capture the physical network's dynamics. Together, these technologies create efficient and robust infrastructure digital twin, which can provide the network controller with holistic infrastructure status, distilled network patterns, and autonomously optimized network decisions.}

\subsection{AI-Native Slice Digital Twin}
{The integration of \acp{rnn}, \ac{llm}, \ac{drl}, and the network simulator is an effective approach to realize slice digital twin-based intelligent network management. RNNs are utilized for generating and processing sequential data within the slice digital twin, particularly useful for analyzing time-series data related to network performance. The \ac{llm} equipped with attention mechanism, is adept at processing and understanding large volumes of textual data. In the context of slice digital twins, the \ac{llm} can analyze user interaction and network traffic data to obtain user satisfaction and network patterns, respectively. The attention mechanism allows for the prioritization of relevant information, making the data analysis more focused and efficient. DRL utilizes sophisticated algorithms where an agent interacts with a simulated network environment to learn and refine slicing policies through trial and error~\cite{s22083031}. This continuous interaction, supported by simulation data from the network simulator, helps in developing robust and efficient network slicing strategies. Overall, AI-native slice digital twin provides effective intelligent network management through data prediction, textual data analysis, and iterative policy learning in a simulated environment.}

\subsection{Workflow}
\begin{figure*}[t]
    \centering  \includegraphics[width=0.85\textwidth]{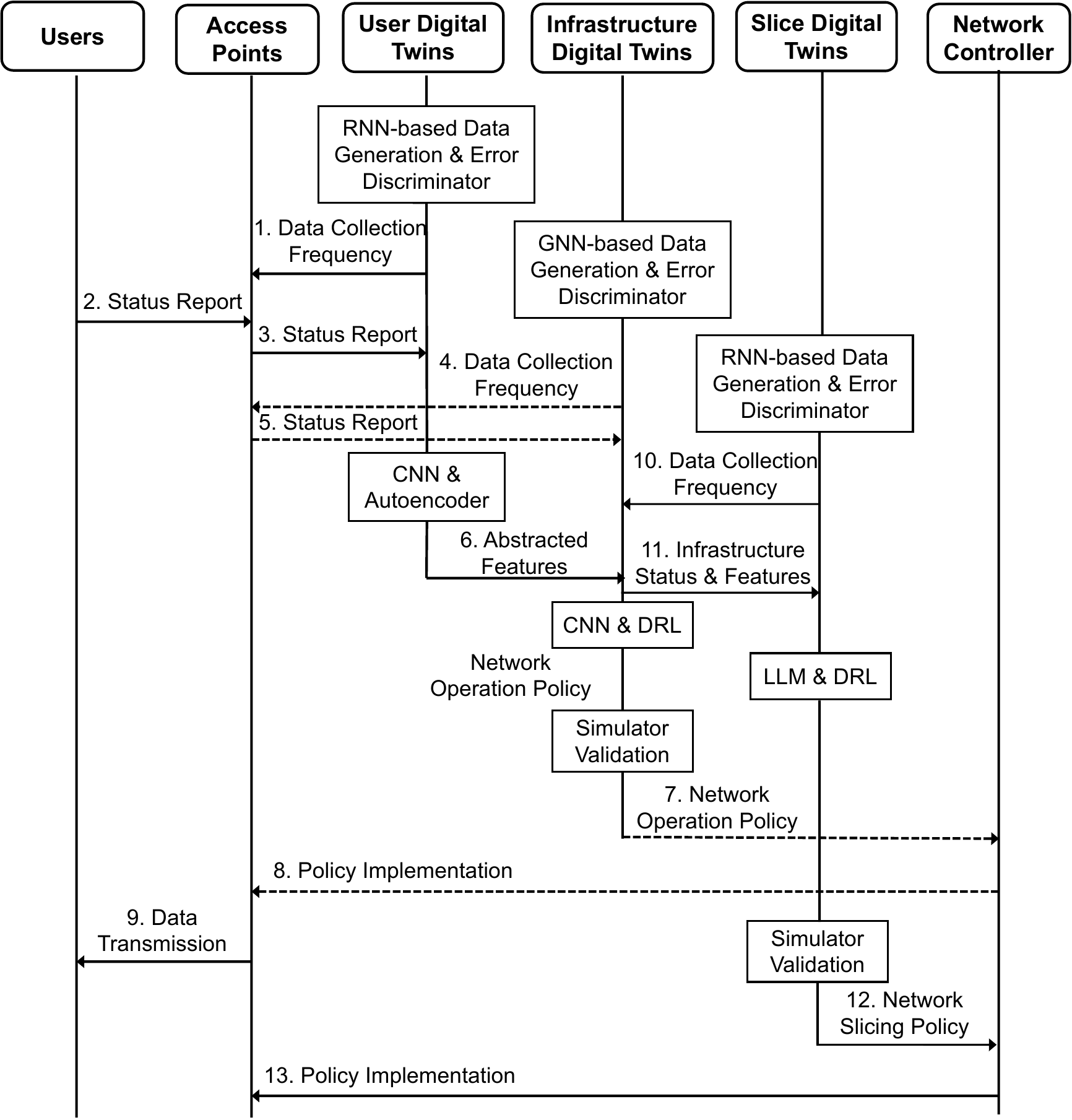}
    \caption{Detailed workflow of the proposed AI-native network digital twin framework.}
    \label{workflow}
\end{figure*}

As shown in Fig.~\ref{workflow}, we present a detailed workflow in the proposed framework. At the beginning, user and infrastructure digital twins employ RNN models to predict future user and network status. If the discrepancy between predicted and actual data surpasses a predefined threshold, user and infrastructure digital twins adjust the data collection frequencies from \acp{ap} to enhance data freshness and precision. Subsequently,  user digital twins utilize CNN and autoencoder models to abstract user features, which are then reported to infrastructure digital twins. Within infrastructure digital twins, CNN models analyze network patterns, feeding into DRL models for crafting network operation policies, which are subsequently validated in an internal infrastructure digital twin simulator to guarantee network robustness. The processed data from  user digital twins and infrastructure digital twins are escalated to slice digital twin, where LLM and DRL models analyze network patterns and user satisfaction to inform network slicing decisions. These decisions undergo validation in an internal slice digital twin simulator to confirm their efficacy and robustness. Ultimately, the infrastructure and slice digital twins forward these refined network operation and slicing decisions to the network controller for implementation.

\section{Challenges and Solutions} \label{sec. Challenges and Solutions}

\subsection{Challenges}
\subsubsection{Efficient Data Collection} 
Constructing network digital twin usually needs to collect massive multimodal data, which places high traffic pressure on mobile communication networks. Multimodal data encompasses various types of information, such as video, audio, and network logs, each requiring different handling, storage, and transmission protocols. The data volume generated by \ac{iot} devices and high-bandwidth applications can be immense. Managing this data influx without causing network congestion and low packet loss is a significant challenge.

\subsubsection{Scalable Data Processing}
Operating and maintaining a network digital twin with low energy consumption is a formidable challenge, primarily due to the need to balance computational intensity with digital twin data processing accuracy. Network digital twin requires continuous and real-time data processing to mirror the physical network accurately. This necessitates substantial computational resources, inherently increasing energy demands. Furthermore, the need for high-fidelity simulations to predict network behavior under different scenarios further escalates the energy requirements. Each simulation might require intensive computational tasks, including data analysis, model updating, and synchronization across multiple digital twin components. Therefore, it is essential to design an efficient digital twin data processing mechanism to achieve energy efficiency. 

\subsubsection{Adaptive Model Update}
Constructing and maintaining a high-fidelity network digital twin, which encompasses user, infrastructure, and slice digital twins, presents significant challenges due to AI model obsolescence caused by data drift. Specifically, a network digital twin usually utilizes a variety of AI models to perform real-time mappings and extract salient network features. However, as network dynamics evolve, changes in user behaviors, device interoperability, and service demands make the data characteristics inherently drift. This drift can significantly degrade the performance of existing AI models, making them obsolete. Therefore, it is critical to timely re-calibrate network digital twin models to align with the network dynamics.




\subsection{Potential Solutions}


\subsubsection{Adaptive Network Digital Twin Model Inference}
To operate and maintain network digital twin with energy efficiency, an adaptive digital twin model inference scheme is paramount. Specifically, this scheme involves employing lightweight AI algorithms for simpler, common networking tasks such as bandwidth allocation or quality of service monitoring, where minimal computational resources are sufficient. For more complex and critical tasks, such as dynamic routing optimization and network anomaly detection, which require deeper analysis and predictive capabilities, more powerful and computationally intensive AI models are activated. This adaptive approach allows the network digital twin to scale its computational demands based on the complexity and criticality of the task, ensuring that the network remains efficient and responsive without excessive energy consumption. Therefore, a versatile and scalable AI model inference scheme should be developed to sustain an energy-efficient network digital twin.
\subsubsection{Dual Error-based Model Update Mechanism}
To effectively counter the challenge posed by AI model obsolescence in the network digital twin, we propose a dual error-based model update mechanism. For data where true labels are accessible, the mean squared error metric is employed to quantify discrepancies between predicted outcomes and actual labels. {Conversely, for data where true labels are unattainable, such as user or network features, we select the entropy of the model’s output probability distribution to measure the error.} This dual-error assessment framework facilitates a flexible model update. The network digital twin model updates are executed through incremental learning, which facilitates model re-calibration without the overhead of retraining from scratch~\cite{6064897}. The proposed error-based model update mechanism ensures that the network digital twin remains robust and accurate in dynamic network environments.

\section{Case Study} \label{sec. Case Study}
\subsection{Considered Scenario}
An adaptive  user digital twin-assisted \ac{msvs} scenario is considered, which consists of multiple base stations, users with \acp{mg}, one edge server equipped with a transcoder for video transcoding, multiple  user digital twins, and a network controller. Users within one base stations are clustered into multiple \acp{mg}, where the same short video streaming is transmitted to one \ac{mg} through multicast transmission. The network operates in a time-slotted manner. In each real-time scheduling slot,  user digital twin can emulate user status and abstract user behavior patterns, which are leveraged by the network controller to make customized communication, computing, and buffer control (3C) management to improve user QoE. The constructed  user digital twin consist of user status data and AI models, i.e.,
\begin{itemize}
    \item  Digital twin data: Two kinds of data, i.e., networking-related data and behavior-related data, are utilized to characterize user status. The former refers to users' channel conditions for estimating users' transmission capabilities. The latter includes users' locations, swipe timestamps, and preferences, which is used to analyze users' mobility trajectories, swipe probability distributions, and video popularity.  User digital twin data originates from two aspects, i.e., realistic data collected from the physical network and predicted data from the \ac{lstm} model. The real-world video streaming dataset\footnote{YouTube 8M dataset: https://research.google.com/youtube8m/index.html} and \texttt{propagationModel} at Matlab are utilized to simulate data collection process from the physical network.
    \item Digital twin models: Two kinds of models, i.e., status prediction model and feature abstraction model, are leveraged to analyze user status. Firstly, based on collected data from the physical network,  user digital twin conducts status prediction by using the \ac{lstm} model to reduce data collection costs. Secondly, the designed autoencoder-based \ac{ddqn} model abstracts users' behavior patterns. Specifically, the autoencoder compresses time-series  user digital twin data into low-dimensional latent features. The \ac{ddqn} analyzes the latent features to determine the appropriate clustering number. Based on the compressed  user digital twin data and clustering number, the K-means++ method can realize fast user clustering while the probability statistics module can analyze the swipe probability distribution.  User digital twin models are updated via incremental learning. The detailed model setting can be found at~\cite{huang2024adaptive}. 

\end{itemize}

The adaptive user digital twin-assisted MSVS operates as follows. In the network operation stage, an appropriate user digital twin feature abstraction model is first activated to conduct model inference for \ac{mg} update. Then, video transcoding and multicast transmission are started based on generative AI-based 3C management to maximize user QoE. A cumulative error detector is used to monitor the accuracy of predicted user status and clustering results. If the cumulative error exceeds a prescribed value, the network enters the  user digital twin model update stage. In this stage,  user digital twin model update is implemented in parallel with video transcoding and multicast transmission. When the  user digital twin model update is completed, the network re-enters the network operation stage.

\subsection{Simulation Results}
\begin{figure}[!t]
  \centering
  \subfloat[Real-time QoE performance in different scheduling slots.]{
  \hspace{2mm} 
\includegraphics[width=7.3cm]{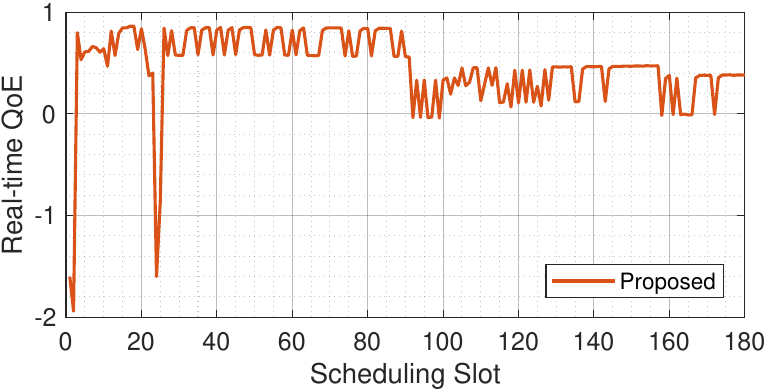}
    \label{fig:4a}
  }
  \newline 
   
  \subfloat[Average QoE performance with respect to the increase of bandwidth.]{
\includegraphics[width=7.4cm]{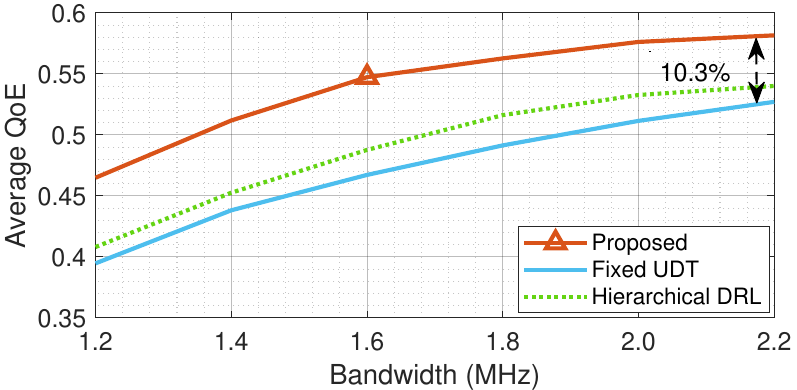}
    \label{fig:4b}
  }
  \caption{The QoE performance comparison.}
  \label{fig:QoE}
\end{figure}

To evaluate the influence of adaptive  user digital twin model inference and update on QoE, the real-time QoE in each scheduling slot is depicted in Fig.~\subref*{fig:4a}. Initially, the QoE value remains low as the  user digital twin model inference conducts multicast grouping and updates the swipe probability distribution, which pauses the video transcoding and transmission process. Around step $20$, QoE decreases a lot because the current  user digital twin model cannot capture new swipe behavior patterns. Then, QoE slightly improves but is disrupted again around step $90$ due to changes in swipe behaviors, necessitating another  user digital twin model update. Overall, the adaptive  user digital twin model inference and update scheme can maintain user QoE at a relatively high and robust level.

For performance comparison, we select two different schemes. Specifically, fixed  user digital twin scheme adopts the fixed  user digital twin model inference version and update frequency with the same 3C management strategy. Hierarchical DRL scheme utilizes the same adaptive  user digital twin model inference and update strategy, but the 3C management strategy relies on a two-layer reward-based DRL algorithm. As shown in Fig.~\subref*{fig:4b}, as bandwidth resources increase, user QoE gradually improves, but the margin of improvement diminishes. This indicates that there is an upper limit to the enhancement of QoE with additional network resources. When the bandwidth increases to 2.2~\myunit{MHz}, the average QoE of proposed scheme is $10.3\%$ higher than that of hierarchical DRL scheme. Overall, the adaptive user digital twin model inference and update scheme with generative AI-based 3C management can achieve a high QoE performance.

\section{Open Research Issues}\label{sec. Open Research Issues}

\subsection{Hierarchical Network Digital Twin Deployment}
{User,  infrastructure, and slice digital twins correspond to different network entities, which are typically deployed at various levels of network nodes: low-level, medium-level, and high-level. However, due to the dynamic and heterogeneous nature of communication networks, it is worthwhile to explore how to flexibly deploy the network digital twins to enhance network performance.} For instance, in scenarios with high mobility, where users frequently switch between base stations, deploying  user digital twins at medium-level nodes can ensure service continuity. Furthermore, to achieve load balancing between macro base stations, infrastructure digital twins can be strategically placed at high-level nodes. Similarly, slice digital twins can be flexibly deployed at medium-level and high-level nodes to maintain high service quality across different service areas.


\subsection{ Hybrid Data-Model Driven Decision-Making}
{Hybrid data-model driven decision-making in the network  digital twin integrates data-driven and model-driven approaches to effectively tackle complex network management problems. Pure data-driven AI methods often struggle to achieve optimal solutions and may lack stability, especially when confronted with unfamiliar scenarios. By decoupling the network management problem, convex sub-problems are solved using model-based techniques for precision and reliability, while complex non-convex sub-problems are addressed with data-driven methods to handle their intricacies. Through iterative interaction between these methods, the network digital twin decision-making progressively converges toward the optimal solution.}


\subsection{Efficient Network Digital Twin Collaboration}
Given the complex tasks that frequently arise in communication networks, such as cooperative sensing and collaborative inference, individual network digital twin struggles to enhance the collaborative capabilities of communication networks. One promising research area could be the integration of advanced multi-agent system algorithms with network digital twin. This approach would allow each network digital twin to act as an independent agent that can negotiate, coordinate, and cooperate with other network digital twins in real-time. Furthermore, a federated learning framework where multiple network digital twins share insights without exchanging raw data can preserve privacy and enable predictive maintenance across different network segments~\cite{10236461}. 


\section{Conclusion} \label{sec:Conclusion}
We have proposed a novel AI-native network digital twin framework to facilitate intelligent network management for 6G. Advanced AI models are utilized to establish network digital twin models and collaborate to complete intricate network analyses and decisions. Network digital twins interact with the physical network to adjust the data collection frequency for timely digital twin model updates. The proposed network digital twin framework can enable efficient and customized management of holistic mobile communication networks to improve user QoE. A case study has been presented, and some open research issues have been discussed to accelerate the pace of network digital twin development.

\section*{Acknowledgment}
This work was supported in part by the Peng Cheng Laboratory Major Key Project under Grants PCL2023AS1-5 and PCL2021A09-B2, in part by the Natural Science Foundation of China under Grant 62201311, and in part by the Young Elite Scientists Sponsorship Program by CAST under Grant 2023QNRC001.
%
\bibliographystyle{IEEEtran}
\bibliography{IEEEabrv,Ref}
%
%

\end{document}